\documentclass[runningheads]{llncs}
\usepackage{bbding}
\usepackage{amsmath}
\usepackage{amssymb}
\usepackage{mathrsfs}
\usepackage{multirow}
\usepackage{booktabs}
\usepackage{color}
\usepackage{ragged2e}
\usepackage[acronym]{glossaries}
\usepackage{graphicx}
\usepackage{caption}
\usepackage{float}
\usepackage{enumitem}
\usepackage{subcaption}
\usepackage[normalem]{ulem}
\useunder{\uline}{\ul}{}
\usepackage{graphicx}
\newcommand{\bm}[1]{\mbox{\boldmath{$#1$}}}
\begin{document}
\title{Dual-Granularity Contrastive Learning for Session-based Recommendation}
\author{Zihan Wang\inst{1} 
\and Gang Wu\inst{1,2(}\Envelope\inst{)} 
\and Haotong Wang\inst{1}}
\institute{School of Computer Science and Engineering, Northeastern University, Shenyang, China \and Key Laboratory of Intelligent Computing in Medical Image, Ministry of Education, Shenyang, China\\
\email{2101816@stu.neu.edu.cn, wugang@mail.neu.edu.cn, 2171931@stu.neu.edu.cn\\}}
\maketitle              
\begin{abstract}
The data encountered by Session-based Recommendation System(SBRS) is typically highly sparse, which also serves as one of the bottlenecks limiting the accuracy of recommendations. 
So Contrastive Learning(CL) is applied in SBRS owing to its capability of improving embedding learning under the condition of sparse data. 
However, existing CL strategies are limited in their ability to enforce finer-grained (e.g., factor-level) comparisons and, as a result, are unable to capture subtle differences between instances. 
More than that, these strategies usually use item or segment dropout as a means of data augmentation which may result in sparser data and thus ineffective self-supervised signals. 
By addressing the two aforementioned limitations, we introduce a novel dual-granularity CL framework. 
Specifically, two extra augmentation views with different granularities are constructed and the embeddings learned by them are compared with those learned from original view to complete the CL tasks. 
At factor-level, we employ Disentangled Representation Learning to obtain finer-grained data, with which we can explore connections of items on latent factor independently and generate factor-level embeddings. 
At item-level, the star graph is deployed as the augmentation method. 
By setting an additional satellite node, non-adjacent nodes can establish additional connections through satellite nodes instead of reducing the connections of the original graph, so data sparsity can be avoided. 
Compare the learned embeddings of these two views with the learned embeddings of the original view to achieve CL at two granularities. 
Finally, the item-level and factor-level embeddings obtained are referenced to generate personalized recommendations for the user. 
The proposed model is validated through extensive experiments on two benchmark datasets, showcasing superior performance compared to existing methods.

\keywords{Session-based Recommendation· Contrastive Learning· Disentangled Representation Learning}
\end{abstract}
\section{Introduction}
\label{sec:intro}
Session-based recommendation (SBR) has gained significant attention recently and it provides recommendations solely based on information from an anonymous session. 

The items that a user interacts with in a short period of time are arranged in chronological order to form a session. 
Session data bears some resemblance to text data in Natural Language Processing (NLP), and as a result, some classic NLP frameworks including Recurrent Neural Network(RNN)\cite{hidasi2015session,quadrana2017personalizing,tan2016improved}, Attention Mechanism\cite{li2017neural,liu2018stamp,luo2020collaborative,xu2019graph} and Graph Neural Network(GNN)\cite{chen2020handling,liu2021case4sr,qiu2019rethinking,wang2020global,wu2019session,yu2020tagnn} have been adapted to SBR. 

GNN-based models have been found to be more effective than other models because they can better model the complex relationships that exist between items. 
However, even with GNN-based models, SBR still faces the challenge of insufficient data to train accurate item embeddings, which can lead to suboptimal recommendation performance. 

Contrastive Learning (CL) in Self-Supervised Learning\cite{wen2021toward,liu2021self,xin2020self} is widely regarded as a solution to the problem of data sparsity. 
The process of CL can be divided into three steps.
Typically, CL models first create a new view through artificial data augmentation based on the original view.
Then learn embeddings from both the original view and the augmentation view. 
Finally, the embeddings learned from the latter are partitioned into positive and negative samples relative to the embeddings learned from the former.
By comparing the differences and similarities between them, models can adjust the embeddings during training following the principle that positive samples are close to each other and negative samples are mutually exclusive. 
As a result, the model can learn more accurate embeddings than with supervised training alone. 
Some researchers have applied CL to the SBR and have achieved promising results\cite{xia2021self,wang2022self,xia2021self2,pan2022collaborative}. 
Nevertheless, we have observed that existing CL models often exhibit two flaws. 
First, existing CL methods are typically limited to coarse-grained comparisons at the item-level and/or session-level, often overlooking finer-grained relationships between instances. 
Second, these methods often rely on item or segment dropout as a form of data augmentation, which can exacerbate data sparsity and consequently, less effective self-supervised signals. 

By addressing the two aforementioned limitations, we introduce a novel dual-granularity CL framework.
Our approach typically involves incorporating two additional augmentation views, one at the item-level and the other at the factor-level, besides the original item-level view. 
The embeddings acquired from two augmentation views are compared with those acquired from the original view to finish CL tasks under two granularities. 

At factor-level, to address the issue of fine-grained factor (e.g. brand and color) labels being often absent in SBR data, we propose the use of Disentangled Representation Learning(DRL) to obtain independent latent factor-level embeddings corresponding to items, which can replace the missing labels. 
So factor-level convolution channels that operate independently can be constructed and then can be compared with the factor-level embeddings converted from the embeddings learned in the original view to finish the factor-level CL. 
At item-level, the star graph has been planted as an augmentation of the original view. 
Star graph has the inclusion of an extra satellite node and non-adjacent nodes can communicate with each other via the satellite node, thereby enabling the acquisition of more information. 
Unlike traditional data augmentation techniques, it avoids further exacerbating data sparsity by promoting nodes to learn more information. 
Likewise, we compared the item-level embedding learned from the original view and star graph to finish item-level CL. 

Moreover, we leverage the learned item-level and factor-level embeddings to model the user's overall and specific latent factor interests, respectively, in order to predict the user's next interaction at two granularities. 
In the end, we propose our model Dual-Granularity Contrastive Learning for Session-based Recommendation(DGCL-GNN) and summarize our main contributions as follows: 
\begin{itemize}
    \item We identify and address two challenges in existing contrastive learning methods for SBR, and propose a dual-granularity contrastive learning framework to improve the model's ability to learn embeddings. 
    \item We innovatively introduce disentangled representation learning and star map augmentation to help us complete two granular comparative learning tasks. 
    \item Eventually, we propose our model DGCL-GNN, and extensive experiments show that the proposed model has achieve statistically significant improvements on benchmark datasets. 
\end{itemize}

This paper is organized as follows. \ref{sec:related work} describes the related work for SBR and CL.  \ref{sec:preliminaries} give formal definitions of the SBR. \ref{sec:methodology} presents the details of our model DGCL-GNN. \ref{sec:Experiments} includes various experiments to demonstrate the effectiveness of our model. Finally, \ref{sec:Conclusion} presents our conclusions and suggestions for future studies. 

\section{Related Work}
\label{sec:related work}
\subsection{Session-based Recommendation}
Session-based recommendation(SBR) has gained considerable research interest in recent years. 
Some researchers have turned to neural network models inspired by NLP, which have shown promise in improving the effectiveness of SBR. 
Recurrent Neural Network(RNN)\cite{medsker2001recurrent} was first noticed because it can capture sequential dependencies in a session, which is essential for SBR. 
However, RNN-based models\cite{hidasi2015session,quadrana2017personalizing,tan2016improved} overlooks the global information that exists within the entire session. 
Some scholars have attempted to incorporate the attention mechanism\cite{vaswani2017attention} to capture global information and proposed some attention-based models\cite{li2017neural,liu2018stamp,luo2020collaborative,xu2019graph}. 
Then Graph Neural Network(GNN)\cite{scarselli2008graph,zhou2020graph} model \cite{chen2020handling,liu2021case4sr,qiu2019rethinking,wang2020global,wu2019session,yu2020tagnn}emerged as a promising solution, showing a significant performance advantage. 
This is due to the GNN's strong ability to capture complex relationships between items, enabling enhanced representation learning capabilities of the model. 
Disen-GNN\cite{li2022disentangled} has introduced DRL, which has become popular in other fields\cite{chartsias2019disentangled,john2018disentangled,tran2017disentangled}, into SBR for the first time. 
This creative approach refines the SBR problem to a fine-grained level, prompting a deeper analysis of the SBR problem.
Although GNN has shown great advantages in SBR, it still faces a significant performance bottleneck, which is the sparsity of the SBR data.

\subsection{Contrastive Learning}
Contrastive learning(CL) is a type of unsupervised learning that aims to learn a representation space where similar samples are mapped to nearby points, while dissimilar samples are mapped to distant points. 
CL has been successfully applied in various fields\cite{dai2017contrastive,khosla2020supervised}, such as computer vision and natural language processing, to improve the quality of learned embeddings and enhance the performance of downstream tasks. 

In the field of recommendation systems like collaborative filtering recommendation and sequential recommendation, CL has also been widely used in recent years\cite{qiu2022contrastive,lin2022improving,chen2022intent}. 
Some studies have also applied CL to the session-based recommendation scenario\cite{xia2021self,wang2022self,xia2021self2,pan2022collaborative}, where it has shown promising results in improving the performance of session-based recommendation models. 

The CL task aims to address a pain point of SBR, as the data in this area is often sparse. 
Therefore, we firmly believe that improving CL is one of the key directions for enhancing SBR's performance in the future.

\section{Preliminaries}
\label{sec:preliminaries}
In this section, we introduce the formal definitions of the session-based recommendation problem. 
Let $\mathcal{I} = \left\{ v_1, ..., v_N\right\}$ denote the set of all items in a dataset and $\bm{N}$ is the number of items. 
Formally speaking, an anonymous session is represented as $\bm{s} = \left\{v_{(s,1)}, ...,v_{(s_n)}\right\}$, where $\bm{n}$ is the total length of the session. 
The task of the SBR models is to analyze the user's interests revealed by interactions in $\bm{s}$ and predict the next item $v_{(s,n+1)}$ that the user most likely interacts with. 
Finally, compare the simulation of user interests with the characteristics of each item to calculate the probability of its potential occurrence $p(v_i|s)$, which represents the score of $v_i$ in this recommendation. 
The Top-K items with the highest scores are selected and recommended to the user. 

\begin{figure*}[htp]
 \centering
 \includegraphics[width=\textwidth]{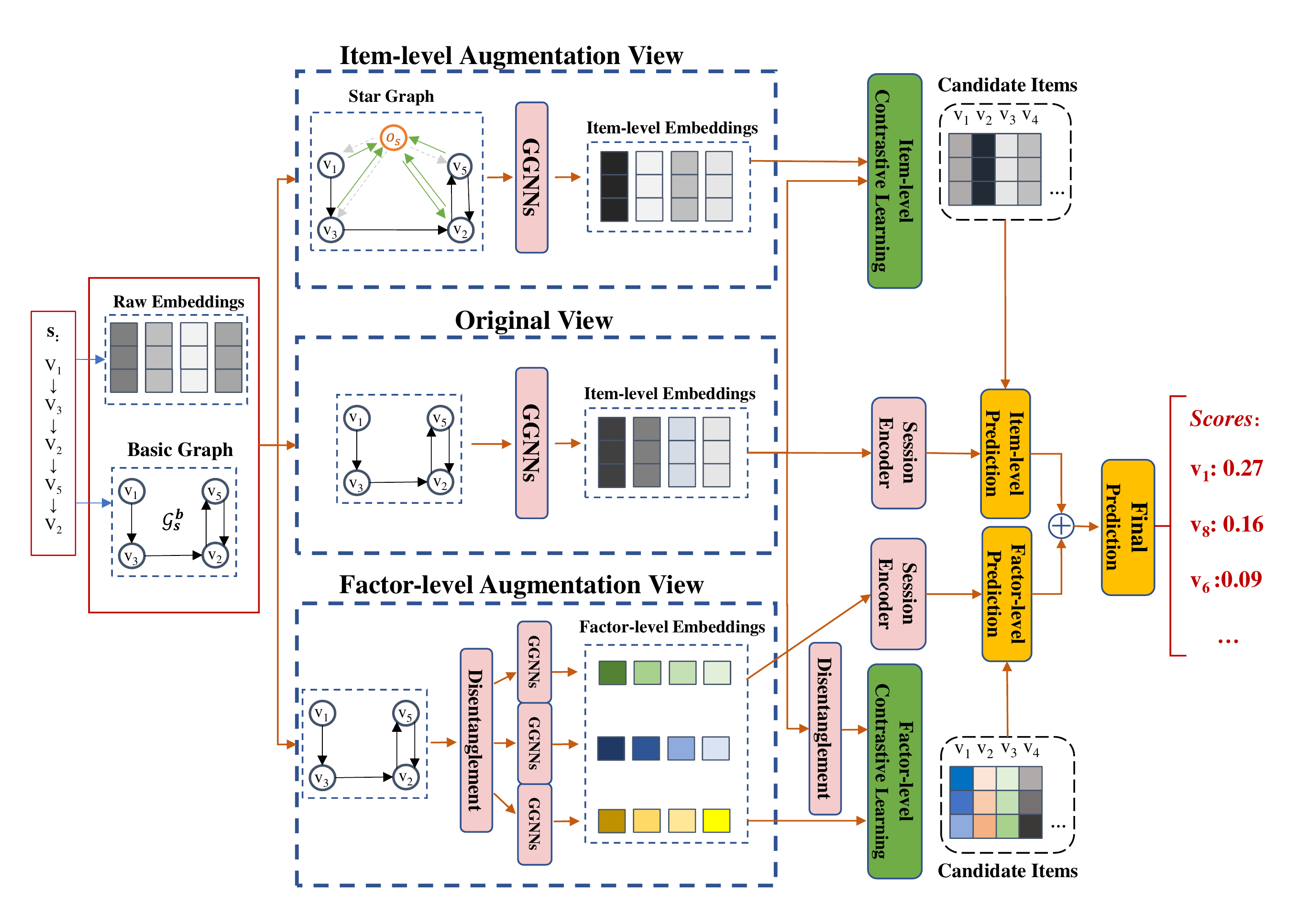}
 \caption{Overview of DGCL-GNN}
 \label{fig:DEISI-Models}
\end{figure*}

\section{Methodology}
\label{sec:methodology}
Next, we describe the framework of our proposed model DGCL-GNN. 
In general, we elaborate on three components of our model mainly including the basic recommendation module, the factor-level CL module and the item-level CL module. 

\subsection{Initialization}
\subsubsection{Original Session Graph}. 
As in the previous work, we use a directed session graph $\mathcal{G}^o_s = <\mathcal{V}^o_s, \mathcal{E}^o_s>$ to store the sequential relationship of each item in the session $\bm{s}$. 
That is, when the next interaction of $v_i$ is $v_j$, we set the $\bm{i}$ row $\bm{j}$ column of the adjacency matrix $\mathcal{A}^o_s$ to 1, which means that there is a pointing relationship. 

\subsubsection{Item-level Embeddings}. 
We first embed items in $\bm{s}$ into the same space to obtain the corresponding item-level embeddings $\bm{e}^s = \left\{ e_{(s,1)},... , e_{(s,n)}\right\}$. 

\subsubsection{Factor-level Embeddings}. 
We adopt DRL to acquire the representations of the independent latent factors underlying the observed data, which corresponds to the fine-grained, factor-level embeddings of the items. 

The input of DRL is the item-level embeddings.
Let $\bm{e}_i\in \mathbb{R}^d$ represents a $\bm{d}$ dimension item-level embedding. 
Then DRL computes $\mathcal{K}$ factor-level embeddings from $\bm{e}_i$ by re-embedding it into corresponding spaces. 
The factor-level embedding $\bm{f}_{(k,i)}$ on factor $k$ is computed with Equation \ref{Equ:DRL}.

\begin{equation}\label{Equ:DRL}
    \bm{f}^k_i = \sigma (W^k_fe_i) + \bm{b}^k_f   (1\leq k \leq \mathcal{K})
\end{equation}

Here, $\bm{W}^f_t \in \mathbb{R}^{d \times d_f}$ and $\bm{b}^f_t \in \mathbb{R}^{d_f}$ are the weight matrix and bias on factor $k$. 
And $d_f = \lfloor \frac{d}{k} \rfloor$ is the dimension of a factor-level embedding.

In order to avoid redundant information between factors, DRL uses the following loss function as the learning objective to generate independent factor-level embeddings.
\begin{equation}\label{Equ:re_emb}
    \mathcal{L}_d = \sum_k^\mathcal{K}\sum_{t\neq k}^\mathcal{K} dCor(f^k, f^t)
\end{equation}
$dCor$ is a formula to measure the correlation between variables in different spaces. 
For other details, please refer to \cite{szekely2007measuring}. 

\subsection{Embedding Learning in Three Channels}
We introduce convolutions in three channels in our model, which aim to learn embeddings for three different views. 

\subsubsection{Original view}. 
The input to this view is the raw item-level embedding $\bm{e}^s = {e_{(s,1)},... , e_{(s,n)}}$ and the raw session adjacency matrix $\mathcal{A}^o_s$. 
In Original view, we use the most commonly used Graph Gated Neural Network(GGNN)\cite{li2015gated} method to update the item-level embedding. 
\begin{equation}
    \begin{split}
    c_{(s,i)}^l = Concat(&W_{in}\mathcal{A}^o_{(s,in)}([x_{(s,1)}^l,...,x_{(s,n)^l}]^\top) + b_{in},\\
                   &W_{out}\mathcal{A}^o_{(s,out)}([x_{(s,1)}^l,...,x_{(s,n)^l}]^\top) + b_{out})
    \end{split}
\end{equation}
\begin{equation}
    z_{(s,i)}^l = \sigma(W_{z}c_{(s,i)}^l + U_{z}x_{(s,i)}^{l-1})
\end{equation}
\begin{equation}
    r_{(s,i)}^l = \sigma(W_{r}c_{(s,i)}^l + U_{r}x_{(s,i)}^{l-1})
\end{equation}
\begin{equation}
    \tilde{x}_{(s,i)}^l = \tanh(W_{h}c_{(s,i)}^l + U_{h}(r_{(s,i)}^l \odot x_{(s,i)}^{l-1}))
\end{equation}
\begin{equation}
    x_{(s,i)}^{l} = (1-z_{(s,i)}^l) \odot h_{i}^{l-1} + z_{(s,i)}^l \odot \tilde{x}_{(s,i)}^l
\end{equation}
Where $\bm{c}_{(s,i)}^l$ is the information that $v_{(s,i)}$ can learn from its adjacent nodes in $\bm{l}$-th layer, ${A}^o_{(s,in}$ and ${A}^o_{(s,out)}$ are the in-degree and out-degree matrices corresponding to the adjacency matrix ${A}^o_s$.  
$x_{(s,i)^l}$ represents the embedding of node $v_{(s,i)}$ at layer $l$ and $x_{(s,i)}^0 = e_{(s,i)}$. 
$z_{(s,i)}^l$ and $r_{(s,i)}^l$ are respectively the update gate and reset gate in the gating mechanism. 
$\bm{\sigma}$ and $\bm{\tanh}$ are activation functions. 
Note that, $W_{in}$, $W_{out}$, $W_{z}$, $W_{r}$, $W_{h}$, $U_{z}$, $U_{r}$, $U_{h} \in \mathbb{R}^{d \times d} $ are all learnable weights and $b_{in}$, $b_{out} \in \mathbb{R}^d$ are biases. 

Finally, we get the updated item-level embedding $\left\{e^o_{(s,1)},...,e^o_{(s,n)}\right\}$.

\subsubsection{Factor-level Augmentation View} 
This view is subdivided into $\mathcal{K}$ independent sub-channels inside, which are respectively responsible for learning the embeddings corresponding to $\mathcal{K}$ hidden factors. 
According to \ref{Equ:re_emb}, measure the distance similarity between factor-level embeddings of each item to generate loss $\mathcal{L}_d$, and finally we hope that each embedding is as independent as possible to reduce the redundant information between sub-channels 

Since items may be similar in one latent factor but not in another, we transform the original session graph $\mathcal{G}^o_s$ to represent the relationship between items on different latent factors. 
We replace the weight in $\mathcal{A}^o_s$ with the corresponding similarity value to generete factor-level adjacency matrixes $\left\{\mathcal{A}^{(f,1)}_s,...., \mathcal{A}^{(f,K)}_s\right\}$. 
$\mathcal{A}^{(f,k)}_s$ is the factor-level adjacency matrix on factor $\bm{k}$.
\begin{equation}
    a^{(f,k)}_{(s,ij)} = \frac{e^k_{(s,i)}e^k_{(s,j)}}{\lvert\lvert e^k_{(s,i)}\rvert \rvert\cdot \lvert\lvert e^k_{(s,j)}\rvert\rvert}
\end{equation}
$a^{(f,k)}_{(s,ij)}$ corresponds to the value of row $\bm{i}$ and column $\bm{j}$ in $\mathcal{A}^{(f,k)}_s$.  

The input of $\bm{k}$-th sub-channel is ,$\left\{e^{k}_{(s,1)}, ... , e^{k}_{(s,n)}\right\}$, the raw factor-level embeddings of items on hidden factor $\bm{k}$ and the factor-level adjacent matrix $\mathcal{A}^{(f,k)}_s$. 
Similar to the original channel, we perform convolution to learn the factor-level embedding based on the adjacency relationships stored in $\mathcal{A}^{(f,k)}_s$ to generate the final output of the sub-channel $\left\{e^{(f,k)}_{(s,1)}, ... , e^{(f,k)}_{(s,n)}\right\}$. 

\subsubsection{Item-level Augmentation View}
 We decide to choose the star graph as the augmentation graph of the original graph at the item-level. 
 In details, we set a satellite node $\bm{O}_s$ for session $\bm{s}$ and the embedding of it $\bm{e}_{(s,O)}$ is set as the average pooling of the items in $\bm{s}$.
\begin{equation}
    \bm{e}_{(s,O)} = \frac{\sum_1^n e_{(s,i)}}{n}
\end{equation}

For the connections between the satellite node and other nodes, we adopt a completely random method that satellite node has an equal probability $\theta$ of pointing or being pointed to with other nodes. 
By implementing the above operations, we can initialize a star graph $\mathcal{G}^*_s$. 
Furthermore, we have the corresponding adjacency matrix $\mathcal{A}^*_s$. 
Then $\bm{e}^s = \left\{e_{(s,1)},... , e_{(s,n)}\right\}$ and $\mathcal{A}^*_s$ are sent into the convolution channel and item-level embeddings are updated as $\left\{e^*_{(s,1)},... , e^*_{(s,n)}\right\}$. 

\subsection{Contrastive Learning}
Once we have learned three embeddings from three views, we can leverage them to accomplish dual-granularity CL tasks. 
Combining two Contrastive Learning (CL) methods can enhance the embedding learning ability of our model. 

\subsubsection{Factor-level Contrastive Learning} 
To begin, we re-embed the output of the original view $\left\{e^o_{(s,1)},...,e^o_{(s,n)}\right\}$ into the corresponding factor-level spaces to get factor-level embeddings $\left\{\left\{e^{(o,1)}_{(s,1)},...,e^{(o,1)}_{(s,n)}\right\},... , \left\{e^{(o,K)}_{(s,1)},...,e^{(o,K)}_{(s,n)}\right\}\right\}$. 
Then we use the factor-level embeddings learned by the factor-level augmentation view $\left\{\left\{e^{(f,1)}_{(s,1)},...,e^{(f,1)}_{(s,n)}\right\},... , \left\{e^{(f,K)}_{(s,1)},...,e^{(f,K)}_{(s,n)}\right\}\right \}$ to compare with the former. 

To perform CL on $\mathcal{K}$ latent factors, we utilize K sub-channels. 
Likewise, we describe the operations on the $\bm{k}$-th channel as an example. 
A standard binary cross-entropy (BCE) loss function has been chosen as our learning objective to measure the difference between the two. 
\begin{equation}
    \mathcal{L}^{(f,k)}_c = - \log~\sigma(H(e^{(o,k)}_{(s,i)}, e^{(f,k)}_{(s,i)}) - \log~\sigma(1-H(e^{(o,k)}_{(s,i)},
    e^{(o,k)}_{(s,j)}) (i \neq j)
\end{equation}
The first half of the comparison involves positive examples, while the second half involves negative examples. and $\bm{H} : \mathbb{R}^{d_f} \times \mathbb{R}^{d_f} \rightarrow \mathbb{R}^{d_f} $is the discriminator function that takes two vectors as the input and then scores the agreement.
between them.

Finally, we aggregate the losses generated by the $K$ channels to obtain the total loss for factor-level CL. 
\begin{equation}
    \mathcal{L}^F_c = \sum^K_{k}{L}^{(f,k)}_c
\end{equation}

\subsubsection{Item-level Contrastive Learning}
We utilize the two item-level embeddings learned from the Original view $\left\{e^o_{(s,1)},...,e^o_{(s,n)}\right\}$ and the item-level augmentation view $\left\{e^*_{(s,1)},... , e^*_{(s,n)}\right\}$ to conduct item-level CL. 
Similarly, we adopt BCE loss as the learning object of item-level CL. 
\begin{equation}
    \mathcal{L}^I_c = - \log~\sigma(H(e^o_{(s,i)}, e^*_{(s,i)}) - \log~\sigma(1-H(e^o_{(s,i)}, e^*_{(s,j)}) (i \neq j)
\end{equation}

The total loss of CL is set as follows: 
\begin{equation}
    \mathcal{L}_c = \alpha * \mathcal{L}^I_c  + (1-\alpha) * \mathcal{L}^F_c
\end{equation} 
$\alpha$ is a hyperparameter controlling the ratio of item-level and factor-level CL losses. 

\subsection{Session Embedding}
The item-level session embedding need to be generated to represent the user's overall preference for items. 
Inspired by Disen-GNN, we also generate factor-level session embeddings in addition, which represent users' interests for specific latent factors. 

\subsubsection{Item-level Session Embedding}
In order to mitigate the potential for misleading predictions, we opt to exclusively employ the embedding derived from the original view to compute the session embedding. 
Although the star graph's enhanced view contains valuable information for contrastive learning, there exists a risk of introducing inaccuracies. 
Therefore, to ensure more reliable predictions, we restrict our utilization to the item-level embeddings obtained from the original view during the session embedding calculation. 
The soft attention is employed as the session encoder. 
\begin{equation}
    \alpha_{(s,i)} = q^\top \sigma(W^{1}_s e^o_{(s,i)}  + W^{2}_s e^o_{(s,n)})
\end{equation}

\begin{equation}
    e^g_s =\sum^n_{i=1}\alpha_{(s,i)} e^o_{(s,i)}
\end{equation}

\begin{equation}
    e^o_s = \bm{W}^{3}_s[e^l_s,e^g_s]
\end{equation}
where $\bm{q} \in \mathbb{R}^d $, $\bm{W}^{1}_s \in  \mathbb{R}^{d \times d}$, $\bm{W}^{2}_s \in  \mathbb{R}^{d \times d}$ and $\bm{W}^{3}_s \in  \mathbb{R}^{d \times 2d}$ are learnable parameters. 
$e^l_s$ and $e^g_s$ represent the session's local and global preferences for factor $t$ respectively. 
And $e^l_s$ is $e^o_{(s,n)}$, the last item's embedding, such setting can make the model pay more attention to the last clicked item, because usually the last item is the most related to the item that the user finally needs. 

\subsubsection{Factor-level Session Embedding}
Similarly, we use the same soft attention to compute the user's preference for $\mathcal{K}$ independent latent factors. 
The process will not be repeated, and finally we concat the user's preferences for $\mathcal{K}$ latent factors as $e^f_s =[e^1_s,... , e^K_s]$. 

\subsection{Prediction and Optimization}
As mentioned earlier, we score candidate items separately based on the item-level and factor-level interests, and then combine the results of them to make the recommendation. 
We use the inner-product as the scoring criterion. 
Note that, in order to calculate the factor-level scores, the item-level embeddings of the candidate items need to be converted into their corresponding factor-level embeddings, and then the inner-product is computed with the corresponding user's factor-level interests. 
The final score of a item is as follows: 
\begin{equation}
    \hat{y_i} = \frac{\hat{y^I_i} + \hat{y^F_i}}{2}
\end{equation}
$\hat{y^I_i}$ and $\hat{y^F_i}$ represent the scoring results of item-level and factor-level respectively. 
$\hat{y_i}$ stores the final scores. 
For the prediction loss, we use the cross-entropy as the loss function, which has been extensively used in the recommendation system: 
\begin{equation}
    \mathcal{L}_p = -\sum^N_{i=1}y_ilog(\bm{\hat{y_i}}) + (1-y_i)log(1-\bm{\hat{y}}_i)
\end{equation}
So, the loss of the whole model consists of three parts: CL, the prediction and DRL. $\beta_1$ and $\beta_2$ are controlling their proportions. 
\begin{equation}
    \mathcal{L} = \mathcal{L}_p + \beta_1 \cdot \mathcal{L}_c + \beta_2 \cdot \mathcal{L}_d
\end{equation}
Adam is adopted as the optimization algorithm to analyze the loss $\mathcal{L}$.

\section{Experiments}
\label{sec:Experiments}
In this section, we introduce the rich experimental content and outline some of the basic experimental settings. 
\subsection{Experiments Settings}
\subsubsection{Datasets}
To verify the effectiveness of DGCL-GNN, we conducted experiments on two commonly used datasets in a session-based recommendation system, \emph{Yoochoose $1/64$}\footnote{http://2015.recsyschallenge.com/challege.html}, and \emph{Diginetica}\footnote{http://cikm2016.cs.iupui.edu/cikm-cup}. 
The statistics of the two datasets are exhibited in Table \ref{datasets}. 
\begin{table}[]
\centering
\caption{Statistical results of datasets}
\label{datasets}
\resizebox{0.5\textwidth}{!}{%
\begin{tabular}{clclcl}
\hline
Statistics  & \multicolumn{2}{c}{Yoochoose1/64} & \multicolumn{2}{c}{Diginetica} \\ \hline
\#interactions  & \multicolumn{2}{c}{557,248} & \multicolumn{2}{c}{982,961} \\
\#training sess  & \multicolumn{2}{c}{369,859} & \multicolumn{2}{c}{719,470} \\
\#test sess  & \multicolumn{2}{c}{55,898} & \multicolumn{2}{c}{60,858} \\
\#items & \multicolumn{2}{c}{16,766} & \multicolumn{2}{c}{43,097} \\
\#avg. length  & \multicolumn{2}{c}{6.16} & \multicolumn{2}{c}{5.12} \\ \hline
\end{tabular}%
}
\end{table}

\subsubsection{Baselines}
To demonstrate the comparative performance of DGCL-GNN, we choose several representative and/or state-of-the-art models. 
They can be categorized into three types: 
(1) \textbf{Non-GNN models}: NARM, GRU4Rec, and STAMP; 
(2) \textbf{Normal GNN models}: SR-GNN and Disen-GNN; 
(3) \textbf{GNN models with CL}: DHCN and COTREC. 
\begin{itemize}
    \item \textbf{GRU4REC}\cite{hidasi2015session} adapts GRU from NLP to SBR.
    As an RNN-based model, it only cares about sequential relationships between items. 
    \item \textbf{NARM}\cite{li2017neural} combined attention mechanism with Gated Recurrent Unit(GRU) to consider both global relationships and sequential relationships to make recommendations. 
    \item \textbf{STAMP}\cite{liu2018stamp} emphasizes the impact of the short time and it designed a special attention mechanism with MLP. 
    \item \textbf{SR-GNN}\cite{wu2019session} introduces GNN to obtain item embeddings by information propagation.
    \item \textbf{Disen-GNN}\cite{li2022disentangled} deployed DRL into SBR to learn the latent factor-level embeddings. Then use the GGNN in each factor channel to learn factor-level session embeddings. 
    \item \textbf{DHCN}\cite{xia2021self} propose a new CL framework, which realizes data augmentation by learning inter-session information. 
    \item \textbf{COTREC}\cite{xia2021self2} is an improved variant of DHCN, it integrated the idea of co-training into CL by adding divergence constraints to DHCN's CL module.
\end{itemize}

\subsubsection{Evaluation Metrics}
Following our baselines, we chose widely used ranking metrics P@$K$(Precise) and M@$K$(Mean Reciprocal Rank) to evaluate the recommendation results where $K$ is 10 or 20. 

\begin{table*}[htbp]
\centering
\caption{Comparing the prediction performance of DGCL-GNN with the baselines. 
The best results in them are highlighted in bold, and the second-best results are underlined. }
\resizebox{0.8\textwidth}{!}{%
\begin{tabular}{c|cccc|cccc}
\hline
\multirow{2}{*}{\textbf{Method}}  & \multicolumn{4}{c|}{\textbf{Yoochoose1/64}} & \multicolumn{4}{c}{\textbf{Diginetica}} \\
  & P@10 & M@10 & P@20 & M@20 & P@10 & M@10 & P@20 & M@20 \\ \hline
NARM  & 0.5920 & 0.2495 & 0.6811 & 0.2855 & 0.3544 & 0.1513 & 0.4970 & 0.1618 \\
GRU4Rec  & 0.5011 & 0.1789 & 0.6063 & 0.2288 & 0.1789 & 0.0730 & 0.2939 & 0.0829 \\
STAMP  & 0.6190 & 0.2583 & 0.6874 & 0.2967 & 0.3291 & 0.1378 & 0.4539 & 0.1429 \\ \hline
SR-GNN  & 0.6197 & 0.2651 & 0.7055 & 0.3094 & 0.3669 & 0.1538 & 0.5059 & 0.1750 \\
Disen-GNN  & 0.6236 & 0.2701 & {\ul 0.7141} & 0.3120 & 0.3981 & 0.1769 & 0.5341 & 0.1879 \\  \hline
DHCN  & 0.6354 & 0.2635 & 0.7078 & 0.3029 & 0.3987 & 0.1753 & 0.5318 & 0.1844 \\
COTREC  & {\ul 0.6242} & {\ul 0.2711} & 0.7113 & {\ul 0.3121} & {\ul 0.4179} & {\ul 0.1812} & {\ul 0.5411} & {\ul 0.1902} \\ \hline
DGCL-GNN   & \textbf{0.6509} & \textbf{0.2889} & \textbf{0.7469} & \textbf{0.3289} & \textbf{0.4305} & \textbf{0.1824} & \textbf{0.5501} & \textbf{0.1911} \\ \hline

\end{tabular}%
}
\label{Tab:Experiments-Overall}
\end{table*}
\subsection{Overall Performance}
Table \ref{Tab:Experiments-Overall} shows the overall performance of DGCL-GNN compared to the baseline models. 
We take the average of 10 runs as the result. 
And we can make the following three observations: 

(1) Compared with RNN-based and Attention-based models, GNN-based models obviously perform better. 
It exhibits the great capability of GNN in learning more accurate embeddings and modeling session data. 

(2) Models trained through CL often exhibit superior performance and demonstrate consistent results across different datasets.  

(3) DGCL-GNN outperforms all the baseline models in all datasets.  
Especially in Nowplaying, there is obvious performance improvement. 

In summary, we can conclude that effective CL can lead to better model optimization, and the enhanced CL in our model has a positive impact on the model's overall performance. 
\begin{figure}[htp]
 \centering
 \includegraphics[width=0.9\linewidth]{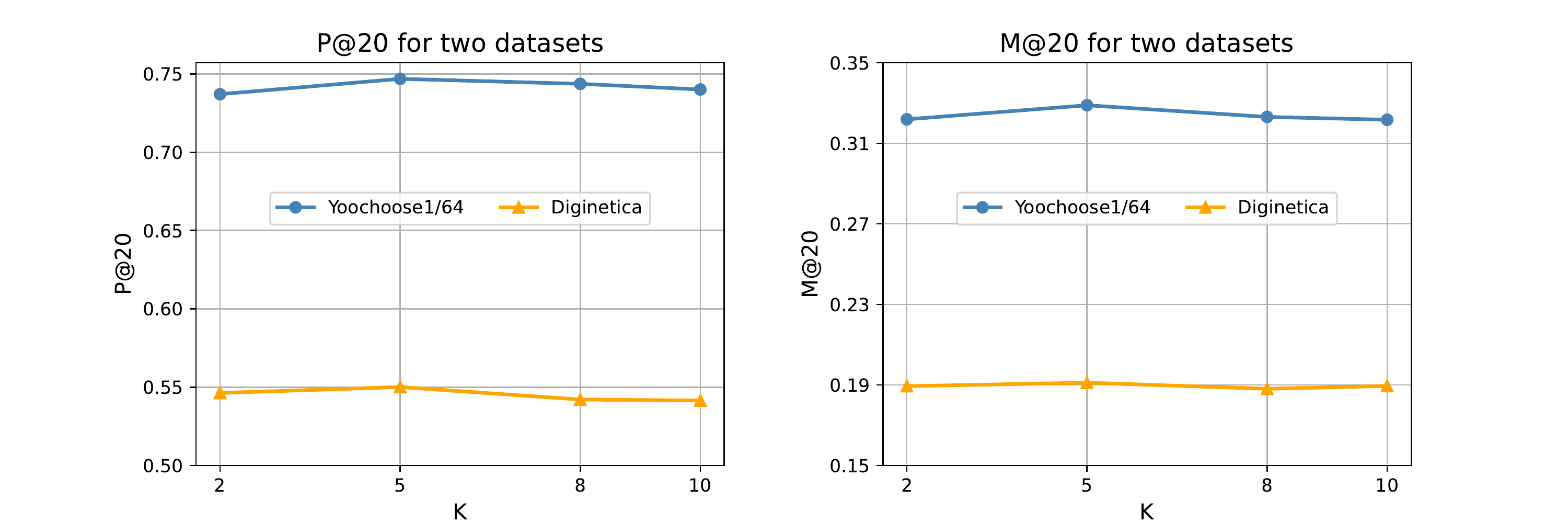}
 \caption{Impact of the hyperparameter $\mathcal{K}$}
 \label{fig:DEISI-Models}
\end{figure}
\subsection{Impact of Hyperparameters} 
The hyperparameter with the greatest impact on model performance is $\mathcal{K}$, which is the number of disentangled hidden factors. 
Having more hidden factors can lead to a finer granularity of factor-level CL, but it can also make it more challenging to train accurate factor-level embeddings. 
Thus, determining the appropriate value for the parameter $\mathcal{K}$ is a crucial consideration.
We set it to 5 on the \emph{Diginetica} and \emph{Yoochoose1/64}.  
The following experimental results also prove the advantages of this setup. 
\subsection{Performance for Different Session Length}
Next, as with most SBR models, we evaluated the performance of our model on both long and short sessions. 
We perform experiments on Yoochoose1/64 dataset and Diginetica dataset. 
We first split the test sets into \emph{long} sessions and \emph{short} sessions. 
\begin{figure}[]
\centering
\begin{subfigure}{\textwidth}
\centering
\includegraphics[width=0.8\textwidth]{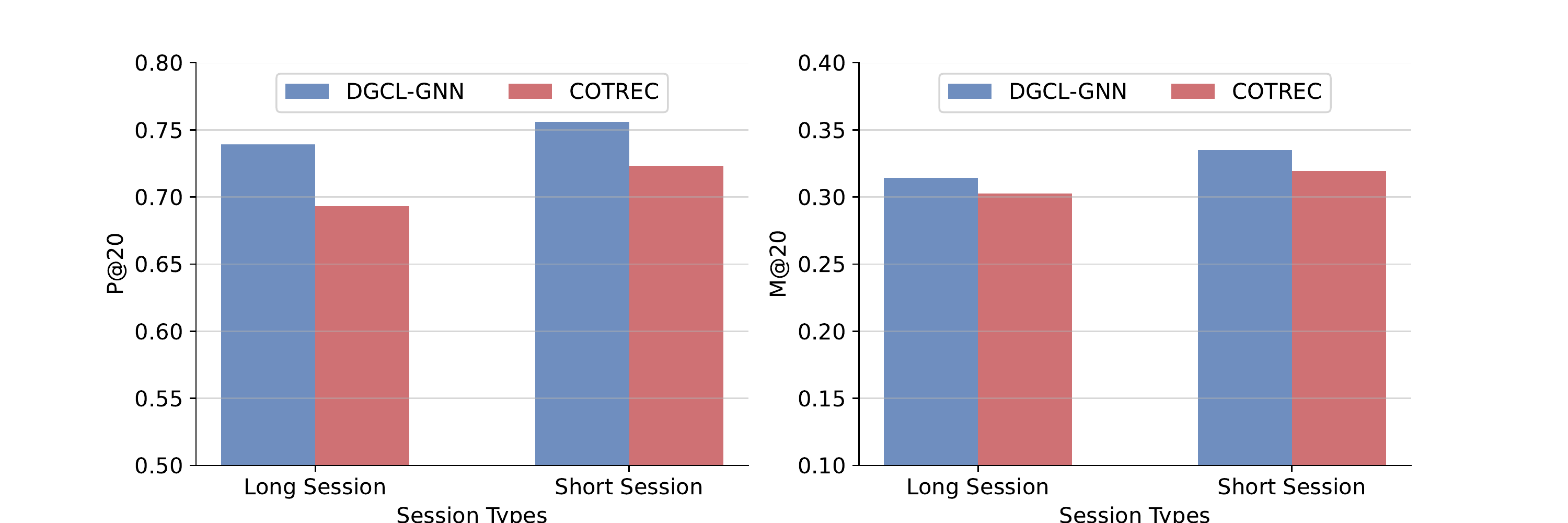}
\caption{P@20 and M@20 on Yoochoose1/64 of short and long sessions}
\label{K_yc}
\end{subfigure}
\begin{subfigure}{\textwidth}
\centering
\includegraphics[width=0.8\textwidth]{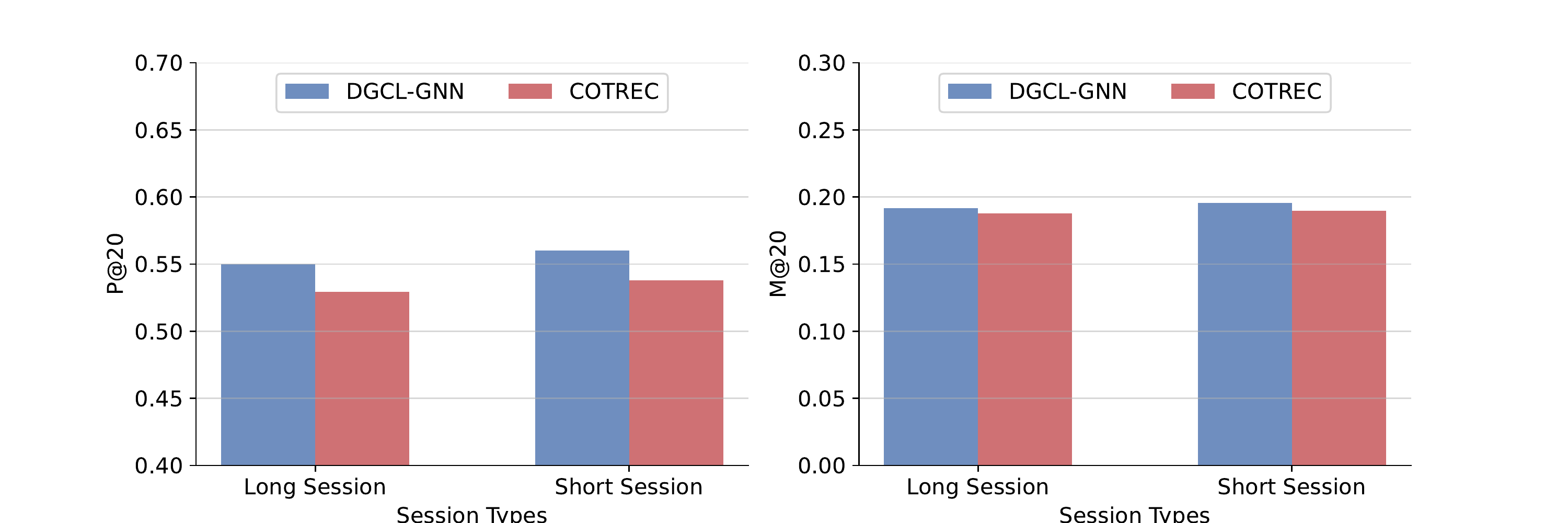}
\caption{P@20 and M@20 on Diginetica of short and long sessions}
\label{K_dg}
\end{subfigure}
\caption{Comparison on different lengths of sessions}
\label{fig:K}
\end{figure}
Similar to \cite{wu2019session}, sessions with a length $\ge$ 5 are defined to be \emph{long} sessions while the others are \emph{short} ones. 
After that, we get two sets of long and short sessions and test the performance of MGCL-GNN and a state-of-the-art baseline, COTREC. 
The experimental results show that DGCL-GNN is superior to the other two models in both long and short sessions. 
\subsection{Ablation Experiments}
In order to verify the effectiveness of the main improvement in our model, the CL module, we have designed three variants of DGCL-GNN.
The first one is DGCL-GNN-FCL, in which we remove the factor-level CL but maintain item-level CL.  
The second one is DGCL-GNN-STAR, we replace the star graph augmentation with the more conventional approach of deleting edges and points as a data augmentation method. 
The third one is DGCL-GNN-FP, We only use item-level user interest to predict the user's next interaction. 
The rest parts of the variants keep consistent with the original DGCL-GNN to ensure the fairness of the ablation experiments. The experimental results prove that any part of our CL module can improve the recommendation performance of DGCL-GNN. 
\begin{table}[]
\centering
\caption{Comparing the performance of DGCL-GNN with its three variants. }
\label{Tab:Ablation}
\resizebox{0.45\textwidth}{!}{%
\begin{tabular}{c|cc|cc}
\hline
\multirow{2}{*}{\textbf{Model}} & \multicolumn{2}{c|}{\textbf{\begin{tabular}[c]{@{}c@{}}Yoochoose\end{tabular}}} & \multicolumn{2}{c}{\textbf{Diginetica}} \\
                   & \textbf{P@20} & \textbf{M@20} & \textbf{P@20} & \textbf{M@20} \\ \hline
\textbf{DGCL}      & \textbf{\textbf{\textbf{\textbf{0.7469}}}}        & \textbf{\textbf{\textbf{0.3289}}}        & \textbf{\textbf{0.5501}}        & \textbf{0.1911}        \\ \hline
\textbf{DGCL-FCL}  & 0.7391        & 0.3201        & 0.5432        & 0.1895        \\ \hline
\textbf{DGCL-STAR} & 0.7415        & 0.3227        & 0.5459        & 0.1902        \\ \hline
\textbf{DGCL-FP}   & 0.7388        & 0.3211        & 0.5369        & 0.1889        \\ \hline
\end{tabular}%
}
\end{table}

\section{Conclusion}
\label{sec:Conclusion}
Our DGCL-GNN model improves the exsiting CL strategy and the dual-granularity CL framework can effectively enhance the model's embedding learning capabilities and thus improve the recommendation accuracy of the model. 
Among them, the additional factor-level contrastive learning can enhance the effect of the original single-grained contrastive learning. The star graph augmentation method also ensures the validity of the self-supervised signal by avoiding destroying the original inter-session graph.  

\subsubsection*{Acknowledgment.} This work was supported by the the National Key Research and Development Program of China (Grant No. 2019YFB1405302).
%

%
%
%
%

\end{document}